\documentclass[pre,showpacs]{revtex4}
\usepackage[dvips]{graphicx}
\usepackage{graphicx}
\begin{document}
\title{Locomotive and reptation motion induced by internal force and friction}
\author{Hidetsugu Sakaguchi and Taisuke Ishihara}
\affiliation{Department of Applied Science for Electronics and Materials,\\
Interdisciplinary Graduate School of Engineering Sciences,\\
Kyushu University, Kasuga, Fukuoka 816-8580, Japan}
\begin{abstract}
We propose a simple mechanical model of locomotion induced by internal force and friction. We first construct a system of two elements as an analog of the bipedal motion. The internal force does not induce a directional motion by itself because of the action-reaction law, but a directional motion becomes possible by the control of the frictional force. The efficiency of these model systems is studied using an analogy to the heat engine.
As a modified version of the two-elements model, we construct a model which exhibits a bipedal motion similar to kinesin's motion of molecular motor. 
Next, we propose a linear chain model and a ladder model as an extension of the original two-element model,. We find a transition from a straight to a snake-like motion in a ladder model by changing the strength of the internal force.  
\end{abstract}
\maketitle
\section{Introduction}
  A uni-directional motion is one of recent topics in nonequilibrium physics. 
The directional motion has been intensively studied, motivated by molecular motors~\cite{rf:1,rf:2,rf:3,rf:4}. Molecular motors such as kinesin and myosin exhibit uni-directional motion along  microtubules and  actin filaments~\cite{rf:5}. Molecular motors are interpreted as energy converters from chemical energy to mechanical energy.   
A ratchet mechanism is considered to be a possible mechanism to generate a directional motion using thermal energy or chemical energy~\cite{rf:6,rf:7}. 
 
Directional motions have been studied also in physical and chemical systems. Linke et al. demonstrated experimentally that liquid drops perform a self-propelled motion when they are placed in contact with hot surfaces~\cite{rf:8}. A sawtooth type asymmetric surface played an important role for the directional motion in their experiments. Chaudhury and Whitesides reported an uphill motion of a liquid drop induced by a surface chemical gradient~\cite{rf:9}.   Sumino et al. found a reciprocal motion of an oil drop on a glass plate induced by oscillatory chemical change of the surface tension~\cite{rf:10}.  
A directional motion was also observed in the active gels where the mechanical motion is coupled with oscillatory chemical reactions~\cite{rf:11,rf:12}. 
In the experiment by Maeda et al., the time-periodic volume change induced by the chemical reaction is transformed to a directional motion on an asymmetric surface. 
We studied a directional motion in a simple model of a self-oscillating elastic filament, motivated by the experiments of the active gels~\cite{rf:13}.
Toyobe et al. performed an experiment of a microscopic particle climbing upwards on a spiral-stair-like potential using a feedback control~\cite{rf:14}. 
    
The self-driving mechanism, especially, the bipedal motion has been one of the main topics in robotics~\cite{rf:15}. The well-controlled walking robots have been developed, where the positions and motions of legs, feet, and many joints are precisely controlled by micro computers.  
However, it is known that there is another simple walking mechanism called passive dynamic walking, where no intricate control is needed~\cite{rf:16}. 
Bipedal walking is realized along a downward slope by a simple mechanical mechanism. As for the locomotion of animals, there are a large amount of literatures in biology.  For example, the reptation motion of snakes was mathematically discussed in \cite{rf:17}. The energetic cost of the snake motion  was estimated~\cite{rf:18}.
Recently, snake-type robots were constructed by combining biology and robotics~\cite{rf:19}. 

In this paper, we propose simple mathematical models based on elementary mechanics.  Directional motion is induced by periodic change of internal force and friction in our model. We study the simple mechanical model systems from a view point of the nonlinear-nonequilibrium physics, hoping that the simple models might be useful to understand directional motions ranging from molecular motors to animal motion and robotics. In \S 2, we study first a simple system composed of two elements, demonstrate a directional motion and discuss the efficiency as an engine. Then, we modify the two-element model to apply to kinesin's motion, and propose  Langevin type equations, because thermal noises are expected to play an important role in molecular motors. 
In \S 3, we extend the original model studied in \S 2 to a linear chain model. 
In \S 4, we generalized the linear chain model to a ladder model where two linear chains are coupled, and we find a dynamical instability from which a snake motion appears. The snake motion is usually considered to be generated by internal muscle forces but our model system suggests that a snake motion might be generated from a dynamical instability, when the body size is sufficiently contracted during a uni-directional motion. 
We consider a sliding friction model and a viscous friction model respectively for a bipedal model and a linear chain model. 
The sliding friction model might be more realistic for the directional motion on the ground.  However, some analytical predictions are possible for the viscous friction models, since the simple system is a linear system, and the viscous friction models might be useful for theoretical understanding of the uni-directional motion.  

\section{Bipedal motion in a system of two elements}
We first study a system composed of two elements as an analogy of the bipedal walking motion. In the bipedal motion, two legs move and stop alternatively. The motion is driven by some internal muscles and stops by the friction with the ground. The frictional force is proportional to the normal force on the ground. The normal force changes periodically by the positional shift of the center of mass of the body. That is, one leg moves forward when the weight of the body is loaded on the other foot, and the roles of two legs are reversed in the next half period. The actual bipedal motion is rather complicated, including intricate controls of several joints and muscles. Here, we consider a very simple mathematical model based on the elementary mechanics: 
\begin{eqnarray}
m\frac{d^2x_1}{dt^2}&=&K(x_2-x_1-a+F\sin\omega t)-W-F_{r1},\nonumber\\
m\frac{d^2x_2}{dt^2}&=&-K(x_2-x_1-a+F\sin\omega t)-W-F_{r2},
\end{eqnarray}
where $x_{1,2}$ is the position of the two elements (analogues of legs), $m$ is the mass, $K$ denotes the spring constant of a spring between the two elements which is an analogue of muscles, $a$ is an average interval between the two elements, and the natural length of the spring is assumed to change in time as $a-F\sin\omega t$ by the internal force. $W$ is the external load. In a climbing motion on a slope of angle $\theta$, the external load is represented as $W=mg\sin\theta$. $F_{r1}, F_{r2}$ are the frictional forces. 
A typical time scale in this model system is $\sqrt{m/K}$, which is proportional to the period of the oscillation of the spring. If the time is rescaled with the typical time $\sqrt{m/K}$, and other parameters are rescaled such as  $W/K\rightarrow W, F_{r1}/K\rightarrow F_{r_1},F_{r2}/K\rightarrow F_{r_2}$ and $(m/K)^{1/2}\omega \rightarrow \omega$, Eq.~(1) is rewritten as 
\begin{eqnarray}
\frac{d^2x_1}{dt^2}&=&(x_2-x_1-a+F\sin\omega t)-W-F_{r1},\nonumber\\
\frac{d^2x_2}{dt^2}&=&-(x_2-x_1-a+F\sin\omega t)-W-F_{r2},
\end{eqnarray}
If the friction and the external load are absent, the center of mass $(x_1+x_2)/2$ does not move if the initial velocities are zero, because of the action-reaction law.
We further assume the sliding friction $F_{r1}$ and $F_{r2}$ of the form:
\begin{eqnarray}
F_{r1}&=&\mu_d(1-\beta\sin\omega t){\rm sgn}(v_1),\nonumber\\
F_{r2}&=&\mu_d(1+\beta\sin\omega t){\rm sgn}(v_2),
\end{eqnarray}
where $\mu_d$ is the coefficient of the sliding friction, and $\beta$ is the amplitude of the periodic variation of the normal force, and $v_1=dx_1/dt,v_2=dx_2/dt$. If the motion stops, the static friction works and the maximum static friction force is expressed as $\mu_s(1\mp \beta\sin\omega t)$. Note that the time-variation of the normal force for the two elements is out of phase, which comes from the shift of the center of mass of the body. 
The average interval $a$ between the two elements is irrelevant in this model. The system of $a=0$ might be better for the bipedal walking, and the model with positive $a$ might correspond to a system composed of fore and hind legs. 

\begin{figure}[t]
\begin{center}
\includegraphics[height=4.cm]{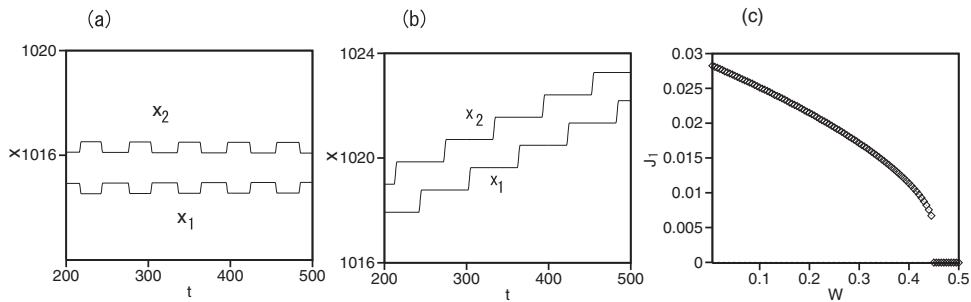}
\end{center}
\caption{(a) Time evolution of $x_1(t)$ and $x_2(t)$ for Eqs.~(2) and (3) at  $a=1.5,F=0.5,W=0,\omega=2\pi/60,\beta=0,\mu_d=0.25$, and $\mu_s=0.6$. (b) Time evolution of $x_1(t)$ and $x_2(t)$ at  $a=1.5,F=0.5,W=0,\omega=2\pi/60,\beta=0.8,\mu_d=0.25$, and $\mu_s=0.6$.  (c) $J_1=\langle dx_1/dt+dx_2/dt\rangle$ vs. $W$  at $a=1.5,F=0.5,\omega=2\pi/60,\beta=0.8,\mu_d=0.25$ and $\mu_s=0.6$.
}
\label{f1}
\end{figure}
At first we show a numerical result for $\beta=0$, that is, the normal force is constant in time.  
Figure 1(a) shows the time evolution of $x_1(t)$ and $x_2(t)$ for $a=1.5,F=0.5,W=0,\omega=2\pi/60,\beta=0,\mu_d=0.25$ and $\mu_s=0.6$. 
Because of the periodic change of the internal force, the interval of $x_2-x_1$ exhibits oscillation. A stick-slip motion of stretch and contraction appears owing to the alternation of the sliding and static frictions. However, the center of the mass $(x_1+x_2)/2$ does not move. 
For $\beta=0.8$, a uni-directional motion appears as shown in Fig.~1(b). 
Sliding motion occurs alternatively for $x_1(t)$ and $x_2(t)$. 
When the spring is contracted, the normal force on the foreleg is sufficiently large and the foreleg cannot move owing to the strong static friction, however, the normal force on the hind leg is relatively weak, and the hind leg can slide forward. On the other hand, when the spring is stretched after a half period, the normal force on the foreleg is weak and the foreleg slides forward, however, the normal force on the hind leg is sufficiently large, and the hind leg cannot move. That is, the hind leg moves forward fixing the position of the foreleg in the first half period, and the foreleg moves forward fixing the position of the hind leg in the second half period.   The internal forces work in the opposite directions for the fore and hind legs because of the action-reaction law of the internal force, but the directional motion is induced owing to the timing between the internal force and the temporal change of the friction force.  This is a physical mechanism of the bipedal motion in our model.  
Even if the external load $W$ is not zero, a uni-directional motion in the $x$-direction is induced, which corresponds to the bipedal walking along an upward slope. Figure 1(c) shows a long time average of the sum of the two velocities: $J_1=\langle dx_1/dt+dx_2/dt\rangle$  as a function of $W$. The other parameter values are set to be $a=1.5,F=0.5,\omega=2\pi/60,\beta=0.8,\mu_d=0.25$ and $\mu_s=0.6$. The average velocity decreases with the external load $W$.  There is a discontinuous transition to the state of $J_1=0$ at $W=0.45$ in this model.   The uni-directional motion is impossible for $W>0.45$. The discontinuous transition occurs because the frictional force is a nonlinear function of the velocity. 

The dynamics is not easily treated mathematically in the nonlinear system. We can propose a linear model with viscous friction:
\begin{eqnarray}
\frac{d^2x_1}{dt^2}&=&(x_2-x_1-a+F\sin\omega t)-W-F_{r1},\nonumber\\
\frac{d^2x_2}{dt^2}&=&-(x_2-x_1-a+F\sin\omega t)-W-F_{r2},\nonumber\\
F_{r1}&=&\mu (1-\beta\sin\omega t)v_1,\nonumber\\
F_{r2}&=&\mu (1+\beta\sin\omega t)v_2,
\end{eqnarray}
where $\mu$ is the coefficient of the viscous friction. 
Figure 2(a) is the time evolution of $x_1(t)$ and $x_2(t)$  for $a=1.5,F=1,W=0,\omega=2\pi/60,\beta=0.5$, and $\mu=20$. 
A forward motion is observed on the average as in Fig.~1(b) even in this simpler model, however, the stick-slip motion does not appear in this model. Although the internal force works in the opposite directions for the two elements, a directional motion appears because of the time-periodic frictional force. This might be interpreted as a kind of ratchet mechanism, in that the reverse motion is effectively suppressed. The direction of the motion is determined by the timing of the variation of the internal force and the frictional force. The energy is injected through the forcing term of $F\sin\omega t$, and is dissipated by the frictional force in this system. That is, our model system is a typical nonequilibrium system. In a molecular motor like the kinesin, the ATP is a source of energy and the energy is dissipated most through the viscous friction with water molecules. 

Since the model equation (4) is a simple linear system, an approximate solution can be obtained by assuming 
\begin{eqnarray}
x_1(t)&=&v_0t+A\sin\omega t+B\cos\omega t,\nonumber\\
x_2(t)&=&a+v_0t-A\sin\omega t-B\cos\omega t.
\end{eqnarray}
Substitution of Eq.~(5) into Eq.~(4) yields
\begin{eqnarray}
-\omega^2 A&=&-2A+\mu \omega B+F+\mu \beta v_0,\nonumber\\
-\omega^2 B&=&-2B-\mu\omega A,\nonumber\\
v_0&=&-W/\mu-\beta\omega B/2,
\end{eqnarray}
if the terms including $\sin(2\omega t)$ and $\cos(2\omega t)$ are neglected. 
The higher harmonics appear through the terms of $-\mu\beta (\sin\omega t) v$ in $F_{r1}$ and $F_{r2}$, and such higher harmonics can be neglected when $\beta$ is sufficiently small. 
The solution of Eq.~(6) is 
\begin{eqnarray}
A&=&\frac{(\omega^2-2)(\beta W-F)}{(\omega^2-2)^2+\mu^2\omega^2(1-\beta^2/2)},\nonumber\\
B&=&\frac{\mu\omega(\beta W-F)}{(\omega^2-2)^2+\mu^2\omega^2(1-\beta^2/2)},\nonumber\\
v_0&=&-W/\mu-\frac{\mu\beta\omega^2(\beta W-F)}{2\{(\omega^2-2)^2+\mu^2\omega^2(1-\beta^2/2)\}}
\end{eqnarray}
A divergence occurs at $\omega=\sqrt{2}$ and $\beta=\sqrt{2}$ in Eq.~(7).
The frequency $\omega=\sqrt{2}$ is the resonance frequency of the linear vibration. In our numerical simulation, $\omega<\sqrt{2}$ is assumed, i.e., the forcing period is sufficiently longer than the period of the proper vibration, and $\beta<1$ is assumed because the normal force becomes negative if $\beta>1$. 
Therefore, the divergence does not occur in our parameter setting.   
For $F=1,W=0,\omega=2\pi/60,\beta=0.5$ and $\mu=20$, $A=0.255, B=-0.269$ and $v_0=0.00703$ are obtained. The dashed curves in Fig.~2(a) represent the approximate solutions (5). Fairly good agreement is seen.  

\begin{figure}[t]
\begin{center}
\includegraphics[height=4.cm]{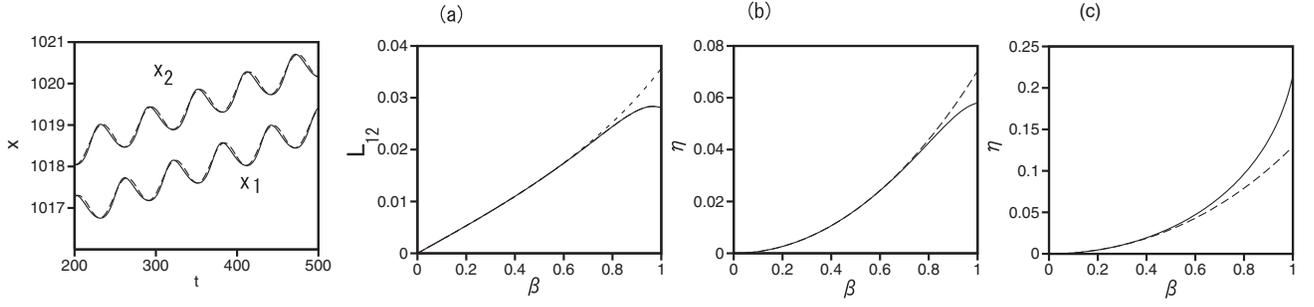}
\end{center}
\caption{(a) Time evolution of $x_1(t)$ and $x_2(t)$ for Eq.~(4) at  $a=1.5,F=1,W=0,\omega=2\pi/60,\beta=0.5$ and $\mu=20$. The dashed curves denote Eq.~(5), whose parameters are given by Eq.~(7). (b) $L_{12}$ (solid line) and $L_{21}$ (dashed line) for the linear model (4) as a function of $\beta$ for $a=1.5,\omega=2\pi/60$ and $\mu=20$. 
The solid and dashed lines are completely overlapped. The dotted line denotes $L_{12}=L_{21}$ by Eq.~(9). (c) Efficiency $\eta$ (solid line) at the maximum power for $a=1.5,\omega=2\pi/60$ and $\mu=20$. The dashed line is the approximation by Eq.~(9). (d) Efficiency $\eta$ (solid line) at the maximum power for $a=1.5,\omega=2\pi/20$ and $\mu=20$. The dashed line is the approximation by Eq.~(9).
}
\label{f2}
\end{figure}

As suggested from Eq.~(7), a uni-directional motion in the $x$-direction is induced also in this viscous friction model, even if $W$ is slightly positive.  When $W$ is positive, the uphill occurs, which implies that this system can perform the "work" in the classical mechanics. If this system is interpreted as an energy converter from mechanical energy to potential energy $W(x_1+x_2)$, the theory of the heat engine can be applied to this system~\cite{rf:20,rf:21}. The efficiency of the heat engine at the maximum power is recently discussed in several thermodynamic systems.  The efficiency at the maximum power is practically useful because the Carnot's maximum efficiency is attained only in the quasi-static process~\cite{rf:22}.   
Van den Broeck  applied Onsager's reciprocal relation to the efficiency of a heat engine at the maximum power. The concept of the efficiency at the maximum power has not yet been studied in detail in deterministic systems. 
In Onsager's theory, two thermodynamic flows $J_1$ and $J_2$  are expressed with the linear functions of two thermodynamic forces $W$ and $F$. In our system, two kinds of flows $J_1=v_1+v_2$ and $J_2=\langle(v_1-v_2)\sin\omega t\rangle$ can be defined, where $\langle\cdots\rangle$ represents a long-time average.  
$J_1$ represents the flow of mass.  The input energy by the internal force per unit time is written as $E=\langle v_1 F\sin\omega t+v_2(-F\sin\omega t)\rangle=FJ_2$. $J_2$ represents a thermodynamic flow with respect to the input energy. 
In the above approximation of Eq.~(5), $J_1=2v_0$ and $J_2=-B\omega$ are obtained, and then, the Onsager type relations are derived in this deterministic system as  
\begin{eqnarray}
J_1&=&L_{11} (-W)+L_{12}F,\nonumber\\
J_2&=&L_{21}(-W)+L_{22}F,
\end{eqnarray}
where 
\begin{eqnarray}
L_{11}&=&1/\mu+\frac{\mu\beta^2\omega^2}{(\omega^2-2)^2+\mu^2\omega^2(1-\beta^2/2)},\nonumber\\  
L_{12}&=&\frac{\mu\beta\omega^2}{(\omega^2-2)^2+\mu^2\omega^2(1-\beta^2/2)},\nonumber\\  
L_{21}&=&\frac{\mu\beta\omega^2}{(\omega^2-2)^2+\mu^2\omega^2(1-\beta^2/2)},\nonumber\\
L_{22}&=&\frac{\mu\omega^2}{(\omega^2-2)^2+\mu^2\omega^2(1-\beta^2/2)}.
\end{eqnarray}
In this approximation, Onsager's reciprocal relation $L_{12}=L_{21}$ is satisfied. The power, that is the work per unit time, is expressed as $P=WJ_1=-L_{11}W^2+L_{12}WF$, and the input energy by the internal force is written as $E=FJ_2$. The dissipated energy by the friction or the waste heat per unit time is expressed as $Q=E-P=L_{11}(-W)^2+(L_{12}+L_{21})(-W)F+L_{22}F^2$. If $L_{21}=L_{12}$, $Q=L_{11}(-W)^2+2L_{12}(-W)F+L_{22}F^2>0$.
The efficiency $\eta$ is expressed as $\eta=P/E$, which is the ratio of the work per unit time over the input energy per unit time. The maximum power is obtained when $\partial P/\partial W=0$, that is, $W=(L_{12}F)/(2L_{11})$. The efficiency at the maximum power is expressed as $\eta=L_{12}^2/(4L_{22}L_{11}-L_{12}L_{21})$. 

Figure 2(b) shows numerically estimated values of $L_{12}$ (solid line) and $L_{21}$ (dashed line) for the linear model (4) as a function of $\beta$ at $a=1.5,\omega=2\pi/60$ and $\mu=20$. The dotted line denotes $L_{12}=L_{21}$ in Eq.~(9). 
The approximation is rather good for $\beta<0.7$ and  Onsager's reciprocal relation is satisfied in the whole range. 
Figure 2(c) shows the efficiency $\eta$ at the maximum power. The dashed line is $\eta=L_{12}^2/(4L_{22}L_{11}-L_{12}L_{21})$ obtained by using Eq.~(9). Good agreement is seen for the efficiency for $\beta<0.7$. For $\beta>0.7$, higher harmonics become important, and the deviation becomes large. The efficiency depends on $\mu$ and $\omega$, i.e., $\eta$ increases with $\mu$ and decreases with $\omega$. Figure 2(d) shows the efficiency $\eta$ at the maximum power as a function of $\beta$ for $a=1.5,\omega=2\pi/20$ and $\mu=20$. The period $2\pi/\omega$ is one third of the case of Fig.~2(c) and the efficiency is fairly increased. 

\begin{figure}[t]
\begin{center}
\includegraphics[height=4.cm]{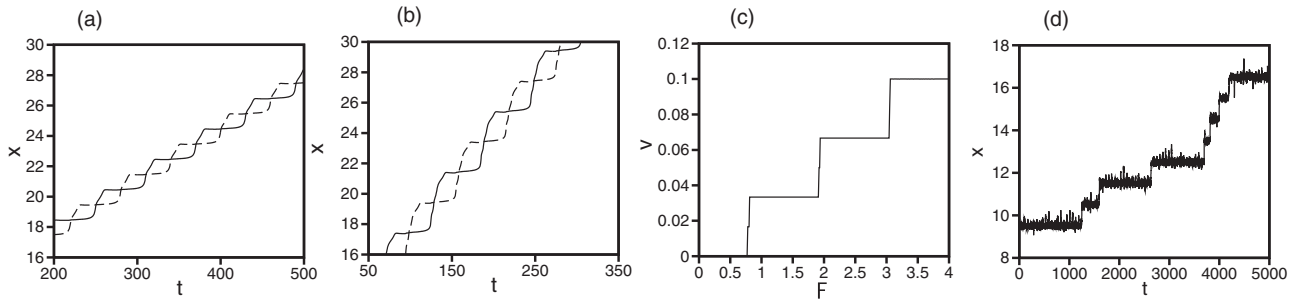}
\end{center}
\caption{(a) Time evolution of $x_1(t)$ (solid curve) and $x_2(t)$ (dashed curve) for $F=1$ at $g=5,\beta=0.9,\omega=2\pi/60, K=2$ and $\mu=10$ in Eq.~(10). (b) Time evolution of $x_1(t)$ (solid curve) and $x_2(t)$ (dashed curve) for $F=2$. 
(c) Average velocity $v$ as a function of $F$ at $g=5,\beta=0.9,\omega=2\pi/60, K=2$ and $\mu=10$. (d) Time evolution of $x_2(t)$ for $F=0.2$ and $T=3$ at $g=5,\beta=0.9,\omega=2\pi/60, K=2$ and $\mu=10$.
}
\label{f3}
\end{figure}
  We can modify the previous model to apply to bipedal motion in molecular motors.  It is known that the kinesin exhibits a bipedal motion along the microtuble. It is characteristics of the system that the microtuble has a spatially-periodic structure of period $l_0$. The one leg of the kinesin becomes docked to the microtuble owing to the bound ATP.  The other leg moves forward by $2l_0$ over the docked leg. At the next half period, the other leg becomes docked and the former leg moves forward by $2l_0$ over the other docked leg.  Stepwise motions occur alternatively for the two legs. Thermal fluctuations play an important role in molecular motors because of the small size.  The effect of thermal noises can be taken in by Langevin ype equations. 
We propose a model for a bipedal motion of the molecular motor along a spatially periodic potential as
\begin{eqnarray}
\frac{d^2x_1}{dt^2}&=&K(x_2-x_1+F\sin\omega t)-\mu \frac{dx_1}{dt}+g(1-\beta\sin\omega t)\sin 2\pi x_1+\xi_1(t),\nonumber\\
\frac{d^2x_2}{dt^2}&=&-K(x_2-x_1+F\sin\omega t)-\mu\frac{dx_2}{dt}+g(1+\beta\sin\omega t)\sin 2\pi x_2+\xi_2(t),
\end{eqnarray}
where the spatial period $l_0$ is assumed to be 1. $\xi_i(t)$ is assumed to be Gaussian white noises which satisfy $\langle \xi_i(t)\xi_j(t^{\prime})\rangle=2T\delta_{i.j}\delta(t-t^{\prime})$ for $i,j=1$ or 2. The parameter $T$ is proportional to the temperature. If $T=0$, Eq.~(10) is reduced to be a deterministic equation. The parameter $a$ is set to be 0, because there is no difference of the fore and hind legs in the kinesin system. The amplitude of the spatially-periodic potential $\cos 2\pi x/(2\pi)$ is assumed to change in time and the timing is out of phase for the two elements. The fact that the attractive force between the kinesin and the microtuble changes in time by the binding of ATP is simply represented by the time-periodic function $1\mp \beta\sin\omega t$ in our model. The two elements tends to be docked near potential minima at $n+1/2$ where $n$ is an integer. On the other hands, the internal force $F\sin\omega t$  tends to enforce exchange of position for the two elements.  
We have first performed numerical simulation of Eq.~(10) for $T=0$ at $g=5,\beta=0.9,\omega=2\pi/60, K=2$, and $\mu=10$.  Figure 3(a) displays time evolutions of $x_1(t)$ (solid curve) and $x_2(t)$ (dashed curve) for $F=1$. Stepwise motion appears for $x_1(t)$ and $x_2(t)$. When $x_1$ stays near a potential  minimum, $x_2$ moves forward by $2l_0=2$.  On the other hand, when $x_2$ stays near a potential minimum, $x_1$ moves forward by $2l_0=2$. The stepwise motion occurs alternately for $x_1$ and $x_2$. 
The average velocity of the stepped motion is $2l_0/(2\pi/\omega)=2/60=0.0333$. Thus, we have succeeded in reproducing qualitatively a bipedal motion of kinesin. When $F$ is smaller than 0.77, such a stepped motion disappears and the two elements are oscillating around a potential minimum. When $F$ is larger than 1.94,   a stepped motion with stepping length $4l_0=4$ appears as shown in Fig.~3(b) for $F=2$. The average velocity is 0.0666 for this stepped motion. 
Figure 3(c) shows the average velocity $v$ as a function of $F$. The average velocity changes stepwise. It is a phenomenon analogous to the frequency locking.

  The effect of thermal noises was studied at $g=5,\beta=0.9,\omega=2\pi/60, K=2,\mu=10, F=0.2$ and $T=3$.  The stepwise forward motion does not occur and the average velocity is zero at $F=0.2$ for $T=0$ as shown in Fig.~3(c). Figure 3(d) shows time evolution of $x_2(t)$ at $T=3$.   A noisy but stepwise forward motion appears in Fig.~3(d). One step is $l_0=1$ in most cases. It implies that the forward motion is induced by the thermal noises. 
Our model is similar to thermal ratchet models studied in [1]-[4], especially similar to a flashing ratchet model~\cite{rf:3}, however,  our model is unique in that the potential is not a sawtooth type but mirror-symmetric. The attractive force to the spatially-periodic filament changes in time and it is out of phase for the two elements, which is essential for the directional motion.

\begin{figure}[t]
\begin{center}
\includegraphics[height=4.cm]{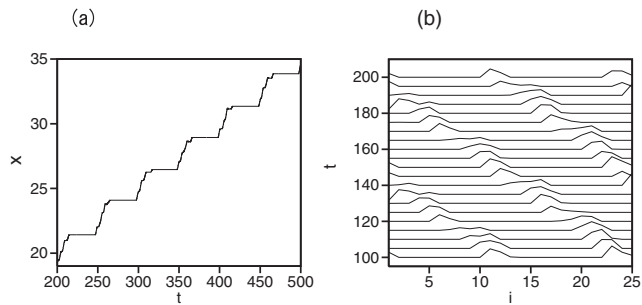}
\end{center}
\caption{(a) Time evolution of $x_i(t)$ for $N=25,i=10$ at $a=1,\theta=\pi/2, F=0.5,\omega=2\pi/50,\beta=0.8,\mu_d=0.25$ and $\mu_s=0.6$ in the sliding friction model (11). (b) Time evolution of the profile $v_i$.
}
\label{f4}
\end{figure}
\begin{figure}[t]
\begin{center}
\includegraphics[height=4.cm]{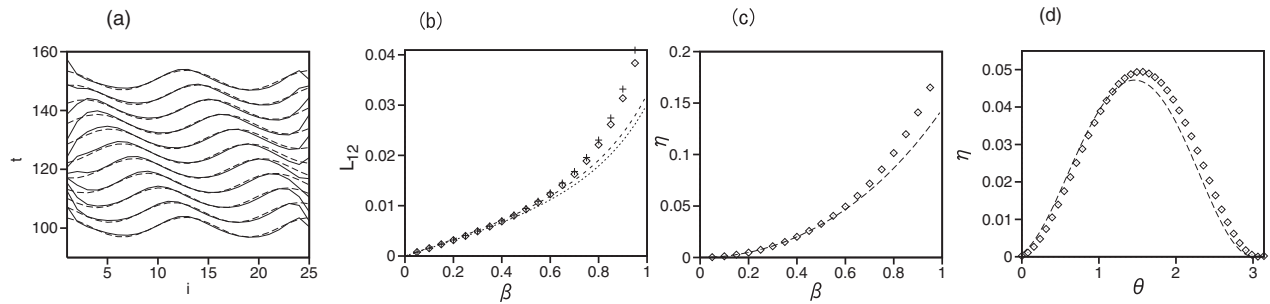}
\end{center}
\caption{(a) Time evolution of the profile $v_i$ for $N=25,W=0,F=1,k=4\pi/N, \theta=\pi/2,a=1,\omega=2\pi/50,\mu=15$ and $\beta=0.2$ in the viscous friction model (12). (b) Onsager's coefficients $L_{12}$ ($+$) and $L_{21}$ (rhombus) as a function of $\beta$ for $N=25,k=4\pi/N, \theta=\pi/2, \omega=2\pi/50$ and $\mu=15$. The dashed lines are the approximation by Eq.~(14). (c) Efficiency $\eta$ as a function of $\beta$ by numerical simulation and the approximation using Eq.~(14). (d) Efficiency $\eta$ as a function of $\theta$ for  $N=25,k=4\pi/N, \omega=2\pi/50,\mu=15$ and $\beta=0.6$  by numerical simulation and the approximation using Eq.~(14).
}
\label{f5}
\end{figure}
\section{Locomotive motion in a linear chain model}
In this section, we consider a one-dimensional chain model as a simple model of an earthworm-type motion.  It is an extension of the two-element models expressed by Eq.~(2) with Eq.~(3), and Eq.~(4).  A sliding friction model is written as
\begin{eqnarray}
\frac{d^2x_i}{dt^2}&=&\{x_{i+1}-x_i-a+F\sin(\omega t+\alpha_i)\}-\{x_i-x_{i-1}-a+F\sin(\omega t+\alpha_{i-1})\}-F_{ri}-W,\nonumber\\
F_{ri}&=&\mu_d\{1-\beta\sin(\omega t+\alpha_i+\theta)\}{\rm sgn}(v_i),
\end{eqnarray}
where $i=1,2,\cdots, N$, $N$ is the total number of elements, $\alpha_i=ki$, and a parameter $\theta$ represents a phase shift between the internal force and the normal force. In this model, the internal force and the normal force are set to propagate like a wave along the linear chain. The parameters $\omega$ and $k$  denote the frequency and the wavenumber of the wave. 
For an earthworm, such a wavy deformation of body segments is actually observed. The normal force on the ground is temporally controlled by the local change of the angle of tiny bristles on the skin. The configuration of the normal force along the long body can be widely changed by a slight change of the anchoring state to the ground, because the normal force tends to be concentrated near locally bristled regions. We have further assumed the free boundary conditions at $i=1$ and $i=N$. 
Figure 4(a) displays time evolution of $x_i(t)$ for $i=10$ at $N=25, a=1,k=4\pi/N, \theta=\pi/2,F=0.5,\omega=2\pi/50,\beta=0.2,\mu_d=0.25$ and $\mu_s=0.6$. A stick-slip motion is observed owing to the alternation of the sliding and the static friction. Figure 4(b) displays time evolution of profile $v_i$ for the same parameter set. The stick-slip motion propagates from the head ($i=N$) toward the tail ($i=1$) like an earthworm. 
 The whole body moves forward successively by the propagation of the locally sliding region from head to tail.  

The viscous friction model is written as
\begin{eqnarray}
\frac{d^2x_i}{dt^2}&=&\{x_{i+1}-x_i-a+F\sin(\omega t+\alpha_i)\}-\{x_i-x_{i-1}-a+F\sin(\omega t+\alpha_{i-1})\}-F_{ri}-W,\nonumber\\
F_{ri}&=&\mu\{1-\beta\sin(\omega t+\alpha_i+\theta)\}v_i.
\end{eqnarray}
Since this viscous friction model is a linear system, it can be solved approximately. If  $x_i(t)=v_0t+A\sin(\omega t+\alpha_i)+B\cos(\omega t+\alpha_i)$ is assumed and the higher harmonics are neglected,   the amplitudes $A$, $B$ and the average velocity $v_0$ are given by 
\begin{eqnarray}
A&=&\frac{\{\mu\beta^2\omega\sin\theta\cos\theta/2-\omega^2+2-2\cos k\}F_1+\mu\omega(1-\beta^2\cos^2\theta/2)F_2}{(\omega^2+2\cos k-2)^2+\mu^2\omega^2-\mu^2\beta^2\omega^2/2},\nonumber\\
B&=&\frac{-\mu\omega(1-\beta^2\sin^2\theta/2)F_1+\{-\mu\beta^2\omega\sin\theta\cos\theta/2-\omega^2+2-2\cos k\}F_2}{(\omega^2+2\cos k-2)^2+\mu^2\omega^2-\mu^2\beta^2\omega^2/2},\nonumber\\
v_0&=&\beta\omega A\sin\theta/2-\beta\omega B\cos\theta/2-W/\mu.
\end{eqnarray}
where $F_1=(1-\cos k)F-W\beta\cos\theta $ and $F_2=F\sin k-W\beta\sin\theta$. 
Figure 5(a) shows time evolution of the profile of $v_i(t)$ for $N=25,W=0,F=1,\theta=\pi/2, k=4\pi/N, a=1,\omega=2\pi/50,\mu=15$ and $\beta=0.2$. 
The deformation propagates smoothly from the head ($i=N$) toward the tail ($i=1$)  and the stick-slip motion does not appear in this model in contrast to the sliding friction model.
The solid curves denote numerical results and the dashed curves express the theoretical result using Eq.~(13). Fairly good agreement is seen except near the two boundaries $i=1$ and $N$. 

We can discuss the efficiency also in this model. Similarly to the two-element model, two types of flows are defined as $J_1=\sum_{i=1}^N\langle v_i\rangle$ and $J_2=\sum_{i=1}^{N-1} \langle(v_i-v_{i+1})\sin(\omega t+\alpha_i)\rangle$. The two types of flows are expanded as $J_1=N\{L_{11}(-W)+L_{12}F\}$ and $J_2=N\{L_{21}(-W)+L_{22}F\}$. The coefficients $L_{11}\sim L_{22}$ are expressed by Eq.~(13) as 
\begin{eqnarray}
L_{11}&=&\frac{1}{\mu}+\frac{\beta^2\mu\omega^2}{2\{(\omega^2+2\cos k-2)^2+\mu^2\omega^2-\mu^2\beta^2\omega^2/2\}},\nonumber\\
L_{12}&=&\frac{\beta\omega[(-\omega^2+2(1-\cos k))\{(1-\cos k)\sin\theta-\sin k\cos\theta\}+\omega\mu\{(1-\cos k)\cos\theta+\sin k\sin\theta\}]}{2\{(\omega^2+2\cos k -2)^2+\mu^2\omega^2-\mu^2\beta^2\omega^2/2\}},\nonumber\\
L_{21}&=&\frac{\beta\omega[(-\omega^2+2(1-\cos k ))\{\sin k\cos\theta-(1-\cos k)\sin\theta\}+\omega\mu\{(1-\cos k)\cos\theta+\sin k\sin\theta\}]}{2\{(\omega^2+2\cos k-2)^2+\mu^2\omega^2-\mu^2\beta^2\omega^2/2\}},\nonumber\\
L_{22}&=&\frac{\omega^2\mu\{\sin^2 k+(1-\cos k)^2\}-\omega^2\mu\beta^2/2\{\sin k \cos\theta -(1-\cos k)\sin\theta\}^2}{2\{(\omega^2+2\cos(k)-2)^2+\mu^2\omega^2-\mu^2\beta^2\omega^2/2\}}
\end{eqnarray}
 Figure 5(b) shows $L_{12}$ ($+$) and $L_{21}$ (rhombus) as a function of $\beta$ for $N=25,k=4\pi/N, \theta=\pi/2, \omega=2\pi/50$ and $\mu=15$. The dashed and dotted curves are respectively $L_{12}$ and $L_{21}$ by the approximation (14). The approximation is good for small $\beta$, but the deviation becomes large for large $\beta$ owing to the higher harmonics. Onsager's reciprocal relation $L_{12}=L_{21}$ is not exactly satisfied in this system. It may be related to a fact that the time-reversal symmetry is not satisfied in the model equation (12). We do not understand well, but Onsager's reciprocal relation might be satisfied in the bipedal model (4) because of an additional symmetry with respect to the exchange of $x_1$ and $x_2$.  
The efficiency $\eta$ at the maximum power is evaluated as $\eta=L^2_{12}/(4L_{22}L_{11}-L_{12}L_{21})$ even in the case of $L_{12}\ne L_{21}$ in this linear system. Figure 5(c) displays numerically obtained values of $\eta$ as a function of $\beta$ and the approximate values using Eq.~(14). The efficiency $\eta$ depends on other parameters. 
Figure 5(d) displays numerically obtained values of $\eta$ as a function of the phase difference $\theta$ and the approximate values using Eq.~(14) for  $N=25,k=4\pi/N, \omega=2\pi/50,\mu=15$ and $\beta=0.6$. The efficiency takes a maximum near $\theta=\pi/2$, and almost 0 near $\theta=0$ and $\pi$. 
This figure implies that the timing between the internal force and the frictional force is very important for the locomotive motion in this model. 

\begin{figure}[t]
\begin{center}
\includegraphics[height=5.cm]{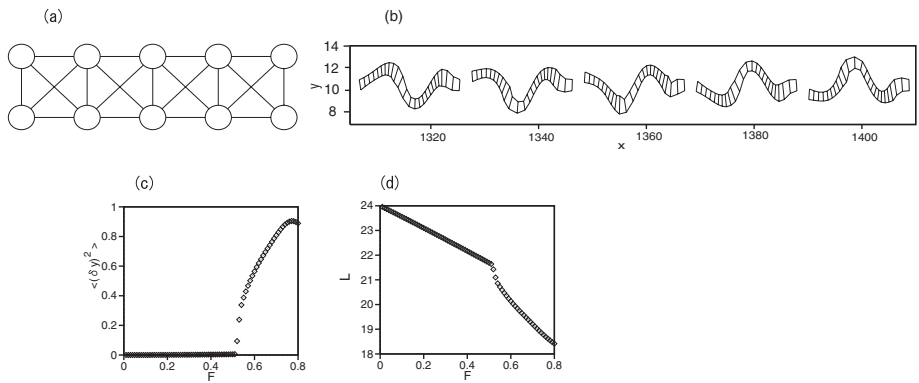}
\end{center}
\caption{(a) Schematic figure of a ladder model. (b) Five snapshot patterns of the ladder model for $N=50,F=0.8,a=1,k=4\pi/25,\theta=\pi/2,\omega=2\pi/50,\beta=0.9$ and $\mu=15$ in the viscous friction model (15) and (16). (c) Mean square $\langle \delta y^2\rangle$ as a function of $F$. (d) Average body length $L$ as a function of $F$. 
}
\label{f6}
\end{figure}
\section{reptation motion in a ladder model}
We can generalize the linear chain model in the previous section to a ladder model where two chain models are laterally coupled as shown in Fig.~6(a) as a very simple model of a snakelike motion. 
In the snakelike motion, the body wriggles in the $y$-direction, when it moves in the $x$-direction on the average. The flex deformation appears along the body in the snakelike motion.  
A linear chain exhibits no resistance against the flex deformation. 
In a ladder model, the flexural rigidity becomes nonzero owing to the finite thickness. This is a reason why we have introduced the ladder model.  

Although the snaking motion was observed in a sliding friction model, we show numerical results of a viscous friction model for the sake of simplicity. 
The viscous friction model is written as 
\begin{eqnarray}
\frac{d^2{\bf x}_i}{dt^2}&=&\sum_j(|{\bf x}_{j}-{\bf x}_i|-a_{ji}){\bf e}_{ji}-F_{ri},\\
F_{ri}&=&\mu \{1-\beta\sin(\omega t+\alpha_i+\theta)\}{\bf v}_i.
\end{eqnarray}
where ${\bf x}_i=(x_i,y_i)$ is the two-dimensional position of the $i$th element, $\alpha_i=ki$ and a parameter $\theta$ denotes a phase difference between the frictional force and the internal force. The elements of $i=1,2,\cdots,N/2$ are located in the lower chain, and the elements of $i=N/2+1,\cdots, N$ are located in the upper chain. 
The summation in Eq.~(15) is taken for the five neighboring sites of $i$, ${\bf e}_{ji}=({\bf x}_j-{\bf x}_i)/|{\bf x}_j-{\bf x}_i|, a_{ji}=a-F\sin(\omega t+\alpha_i)$ for the right site,  $a_{ji}=a-F\sin(\omega t+\alpha_{i-1})$ for the left site, $a_{ji}=a$ for the opposite site, $a_{ji}=\sqrt{a^2+\{a-F\sin(\omega t+\alpha_i)\}^2}$ for the diagonally right site, and $a_{ji}=\sqrt{a^2+\{a-F\sin(\omega t+\alpha_{i-1})\}^2}$ for the diagonally left site. 
Figure 6(b) shows five snapshot patterns of the ladder model for $N=50,F=0.8,a=1,k=4\pi/25,\omega=2\pi/50,\beta=0.9$ and $\mu=15$.
The time interval between the two snapshots is 800.  In this simulation, the $y$ coordinate $y_{N/2}(t)$ and $y_{N}(t)$ at the right boundary are respectively fixed to 10 and 11.  This boundary conditions induce a uni-directional motion in the $x$-direction on the average. If the free boundary conditions are imposed at the right boundary, the direction of motion changes in time. The ladder moves forward in the $x$-direction and it is deformed in the $y$-direction. As a result, a snakelike motion appears. 
We have calculated the mean square $\langle \delta y^2\rangle=\langle y_i^2\rangle-\langle y_i\rangle^2$ to quantify the deformation in the $y$-direction. The average with respect to $i$ and the long-time average are taken for the calculation of $\langle \delta y^2\rangle$. Figure 6(c) shows $\langle \delta y^2\rangle$ 
as a function of $F$.
The mean square $\langle \delta y^2\rangle$ changes continuously from 0 to nonzero values at $F=F_c\sim 0.52$, which implies the appearance of the snaking pattern.  The ladder is straight and exhibits an earthwormlike linear motion for $F<F_c$. The straight state becomes unstable and the snake pattern appears for $F>F_c$. Figure 6(d) shows the average body length $L=(x_{N/2}+x_{N}-x_1-x_{N/2+1})/2$ as a function of $F$. The average body length $L$ decreases with $F$ monotonously. The snaking pattern appears for $F>F_c$, when the body length becomes sufficiently short.   

When an elastic rod is compressed, a buckling instability occurs if the compressive force $F$ is larger than a critical value $F_c$.~\cite{rf:23,rf:24}.  In the classic Euler buckling of an elastic rod, a straight rod becomes unstable and a characteristic deformation of one-hump appears for $F>F_c$. The wavenumber of the deformation is $q=\pi/L$ where $L$ is the system size. 
The instability shown in Fig.~6(c) is interpreted as a kind of the buckling instability, however, a wriggle structure with  many humps is characteristic in our system.  

\begin{figure}[t]
\begin{center}
\includegraphics[height=6.cm]{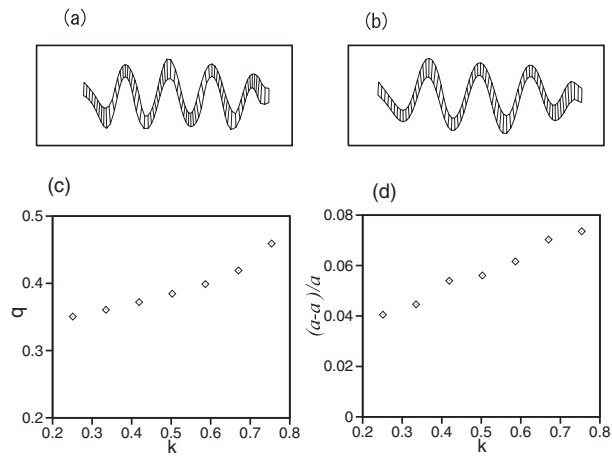}
\end{center}
\caption{(a) A snapshot pattern of the ladder model for $N=150,F=0.69,a=1,k=4\pi/25,\theta=\pi/2,\omega=2\pi/50,\beta=0.9$ and $\mu=15$ in the viscous friction model. (b)  A snapshot pattern of the ladder model for $k=2\pi/25$. The other parameters are the same as (a). (c) Wavenumber $q$ of the snaking pattern as a function of $k$ for $F=0.4$ and $N=150$. (d) Critical value  of the compression rate $(a-\bar{a})/a$ as a function of $k$ for $N=150$. 
 }
\label{f7}
\end{figure}
We have numerically studied the wavenumber $q$ of the snaking pattern in a larger system with $N=150$. 
Figure 7(a) shows a snapshot pattern for $k=4\pi/25$ and $F=0.69$. It is natural that the wriggle number of the pattern is larger in this larger system compared to the system of $N=50$ shown in Fig.~6(b). Figure 7(b) shows a similar snapshot pattern for $k=2\pi/25$ and $F=0.69$. 
The scale of the $x$ and $y$ coordinates is the same in Figs.~7(a) and (b).
The wriggle number of the pattern is larger at $k=4\pi/25$ than at $k=2\pi/25$. The average body size $L=(x_{N/2}+x_{N}-x_1-x_{N/2+1})/2$ is more contracted at $k=4\pi/25$ than at $k=2\pi/25$.  
We have studied the instability more in detail by changing the parameter $k$. 
Figure 7(c) denotes the average wavenumber $q$ of the snaking pattern as a function of $k$ for $F=0.4$. Here, the average wavenumber $q$ was numerically evaluated from the wavelength between the successive two peaks of the snaking patterns. The average wavenumber $q$ is much larger than the value $\pi/L_0=0.042$ of the one-hump structure, where $L_0=N/2-1$.  And, the average wavenumber $q$ is definitely different from $k$, but increases with the parameter $k$. 
We have calculated the mean square $\langle \delta y^2\rangle$ and the average body length $L=(x_{N/2}+x_{N}-x_1-x_{N/2+1})/2$ as a function of $F$ at various $k$'s for $N=150$ to find the critical points. 
The bifurcation occurs when the compression rate $(a-\bar{a})/a$ is larger than a critical value.  Here, $\bar{a}=L/(N/2-1)$ denotes the average interval between two neighboring elements in the $x$-direction, and $a=1$ is the natural interval. 
Figure~7(d) shows the numerically obtained compression rate at the bifurcation point as a function of $k$ for $N=150$.  The critical compression rate tends to increase  with $k$. 
We do not sufficiently understand why a snake pattern with $q\gg \pi/L$ appears in the ladder model. We neither understand the $k$-dependence of $q$ and  the compression rate shown in Figs.~7(c) and (d). The analysis based on the theory of elasticity is left to future study. 

\section{Summary}
We have constructed several simple mechanical models which exhibit a uni-directional motion induced by internal forces and frictions. 
The model composed of two elements was first constructed for the bipedal motion. The energy is injected through the periodic change of the natural length of the spring, which plays a role of a kind of a muscle, and dissipated via the friction. This 
is  a typical nonequilibrium system, although thermal fluctuations are not considered. 
The bipedal motion is caused by a kind of ratchet mechanism in which the reverse motion is suppressed by temporally enforced frictions.  
The sliding friction models might be more plausible, but the viscous friction models can be solved approximately. When the locomotive motion proceeds along an upward slope, the input energy is transformed to the potential energy and a part of the input energy is dissipated by the friction. The efficiency at the maximum power is evaluated in these deterministic systems and it is compared to the theory of Van den Broeck. 
We have also proposed a modified model for the bipedal motion along a spatially periodic potential as a simple model of kinesin's motion.  The effect of thermal noises was studied using the Langevin type equations.  

The bipedal models have been extended to a linear chain model and a ladder model.  In the viscous friction model, we have found that Onsager's reciprocal relation is not always realized even if it is a linear system. In the ladder model, we have found a transition from the earthworm motion to the reptation motion when the forcing amplitude is increased. The transition occurs when the ladder is sufficiently compressed by the internal force and the friction. 

Our model systems are very simple models to understand directional motions theoretically. More realistic models including many degrees of freedom or three-dimensional deformations need to be developed to study specific systems such as molecular motors and real snakes.


\begin{thebibliography}{99}
\bibitem{rf:1} F.~Julicher, A.~Ajdari, and J.~Prost, Rev. Mod. Phys.{\bf 69},1269 (1997).
\bibitem{rf:2} M.~O.~Magnasco, Phys. Rev. Lett. {\bf 71}, 1477 (1993).
\bibitem{rf:3} R.~D.~Astumian and M.~Bier, Phys. Rev. Lett. {\bf 72}, 1766 (1994).
\bibitem{rf:4} H.~Sakaguchi, J. Phys. Soc. Jpn. {\bf 67}, 709 (1998).
\bibitem{rf:5} R.~D.~Vale and R.~A.~Milligan, Science {\bf 288}, 88 (2000).
\bibitem{rf:6} R.~D.~Vale and F.~Oosawa, Adv, Biophys. {\bf 26}, 97 (1990).
\bibitem{rf:7} A.~Ishijima, T.~Doi, K.~Sakurada, and T.~Yanagida, Nature {\bf 352}, 7301 (1991). 
\bibitem{rf:8} H.~Linke, B.~J.~Alem\'an, L.~D.~Melling, M.~J.~Taormina, M.~J.~Francis, C.~C.~Dow-Hygelund, V.~Narayanan, R.~P.~Taylor and A.~Stout, Phys. Rev. Lett. {\bf 96}, 154502 (2006).
\bibitem{rf:9} M.~K.~Chaudhury and G.~M.~Whitesides, Science {\bf 256}, 1539 (1992).
\bibitem{rf:10} Y.~Sumino, N.~Magome, T.~Hamada, and K.~Yoshikawa, Phys. Rev. Lett. {\bf 94}, 068301 (2005). 
\bibitem{rf:11} R.~Yoshida, T.~Takahasi, T.~Yamaguchi and H.~Ichijo, J. Am. Chem. Soc. {\bf 118}, 5134 (2000).
\bibitem{rf:12} S.~Maeda, Y.~Hara, R.~Yoshida and S.~Hashimoto, Int. J. Mol.Sci. {\bf 11}, 52 (2010).
\bibitem{rf:13} H.~Sakaguchi, Phys. Rev. E {\bf 79}, 026216 (2009).
\bibitem{rf:14} S.~Toyabe, T.~Sagawa, M.~Ueda, E.~Muneyuki, and M.~Sano, Nature Physics {\bf 6}, 988 (2010).
\bibitem{rf:15} e.g., J.~Yamaguchi, E.~Soga, S.~Inoue, and A.~Takahashi, Proc. of the 1999 IEEE Int. Conf. Robotics and Automation,  pp 368,1999. 
\bibitem{rf:16} T.~MacGeer, Int. J. Robotic Research {\bf 9}, 62 (1990).
\bibitem{rf:17} e.g.,N.~Rashevsky, {\it Mathematical Biophysics} (Diver Publication Inc., 1960).  
\bibitem{rf:18} M.~Walton, B.~C.~Jayne, A.~F.~Bennett, Science {\bf 249}, 524 (1990).
\bibitem{rf:19} W.~Watanabe, T.~Sato and A.~Ishiguro, Proc. of 2009 IEEE/RSJ Int. Conf. Inteligeny Robots and Systems, 2421 (2009).
\bibitem{rf:20} C.~Van den Broeck, Phys. Rev. Lett. {\bf 95}, 190602 (2005).
\bibitem{rf:21} A.~Gomez-Marin and J.~M.~Sancho, Phys. Rev. E {\bf 74}, 062102 (2006).
\bibitem{rf:22} F.~Curzon and B.~Ahlborn, Am. J. Phys. {\bf 43}, 22 (1975).
\bibitem{rf:23} e.g. L.~D.~Landau and E.~M.~Lifshitz, {\it Theory of Elasticity} (Pergamon, New York, 1986).
\bibitem{rf:24} J.~R.~Gladden, N.~Z.~Handzy, A.~Belmonte, and E.~Villermaux, Phys. Rev. Lett. {\bf 94}, 035503 (2005). 
\end{thebibliography}
\end{document}